\documentclass[12pt,a4paper,final]{iopart}
\usepackage{iopams}
\usepackage{graphicx}
\usepackage[breaklinks=true,colorlinks=true,linkcolor=blue,urlcolor=blue,citecolor=blue]{hyperref}
\usepackage{subfigure}
\usepackage{mathrsfs}

\begin{document}

\title{An efficient Multiple Scattering method based on partitioning of scattering
matrix by angular momentum and approximations of matrix elements}

\author{Junqing Xu}
\address{National Synchrotron Radiation Laboratory, University of Science and
 Technology of China, Hefei, Anhui, 230026, China}

\author{Keisuke Hatada$^{1,2,3}$}
\address{$^1$D\'{e}partement Mat\'{e}riaux Nanosciences, Institut de Physique de Rennes
 UMR UR1-CNRS 6251, Universit\'{e} de Rennes 1, F-35042 Rennes Cedex, France}
\address{$^2$Physics Division, School of Science and Technology, Universit\`a di Camerino,
 via Madonna delle Carceri 9, I-62032 Camerino (MC), Italy}
\address{$^3$INFN Laboratori Nazionali di Frascati, Via E Fermi 40, c.p. 13, I-00044
 Frascati, Italy}
\eads{\mailto{keisuke.hatada@univ-rennes1.fr}}

\author{Didier S\'ebilleau}
\address{$^1$D\'{e}partement Mat\'{e}riaux Nanosciences, Institut de Physique de Rennes
 UMR UR1-CNRS 6251, Universit\'{e} de Rennes 1, F-35042 Rennes Cedex, France}

\author[cor1]{Li Song}
\address{National Synchrotron Radiation Laboratory, University of Science and
 Technology of China, Hefei, Anhui, 230026, China}
\eads{\mailto{song2012@ustc.edu.cn}}

\begin{abstract}
We present a numerically efficient and accurate Multiple Scattering formalism,
which is a generalization of the Multiple Scattering method with a truncated
basis set [X. -G. Zhang and W. H. Butler, Phys. Rev. B 46,7433 (1992)].
Compared to the latter method, we keep the phase shifts of high angular momenta
but apply approximations in the matrix elements of the scattering matrix ($I-tg$), which is the
subtraction of the unit matrix and the product of transition operator matrix
and structure constant matrix. We have discussed the detailed behaviour of our formalism for
some different types of calculations, where not full information of Green's function
is needed. We apply our formalism to study density of states of
fcc Cu and silicon and C K-edge X-ray absorption spectra of graphene, in order
to check the efficiency and accuracy of our formalism. It is found that compared to
Zhang's method, the accuracy is greatly improved by our method.

\end{abstract}

\section{Introduction}

Multiple Scattering (MS) theory was proposed originally by Korringa and by Kohn
and Rostoker (KKR) as a convenient method for calculating the electronic
structure of solids \cite{KKR1,KKR2} and was later extended to polyatomic
molecules by Slater and Johnson \cite{Slater}. It has been widely applied to
the computation of the electronic structure
\cite{Rev_MS_Ebert,stocks1978complete,zeller1995theory} and spectroscopies
\cite{Didier,feff,ZYW} of various systems, such as crystals, molecules,
surfaces, alloys, as well as systems with defects and adsorbates.\par

In MS, wavefunctions are expanded in a spherical-wave basis, which is the local
numerical solution of the Schrodinger equation (SE). In practice, the size of
basis sets is truncated to a particular orbital angular momentum - $l_{max}$.
For some studies where accuracy is not very crucial or angular moment
convergence is very fast, $l_{max}$ can be reasonably chosen to a small value,
e.g., 2, 3 or 4. However, in some applications or if we need higher accuracy,
$l_{max}$ will be not that small, e.g., 6 to 8 or even higher. The MS
computation for one energy point consists of three steps: (i) Solving the
single-site SE to obtain the transition operator matrix $t^i_{LL'}$ and
constructing the structure constant matrix $g^{ij}_{LL'}$, where $i$ represents
a scattering site and $L\equiv(l,m)$ is the combination of orbital and magnetic
angular momenta. The definition of $t^i_{LL'}$ is given in Eq. \ref{eq:t} of
Sec. \ref{sec_GF}. (ii) Inverting the scattering matrix which is defined in our
formalism as $I-tg$, where $I$ represents the unit matrix. (iii) Computing the
Green's function and the studied properties.\par
Nowadays, when deal with nanostructured systems, molecular adsorption on
surface, systems with impurities and some complicated materials, e.g., lithium
ion battery, people have to build huge unitcells or cluster models for
computations. Therefore, it becomes important to reduce the computation time
without much loss of accuracy. In MS calculations, since the computation time
of matrix inversion is proportional to the cube of the dimension of the matrix,
if the system is large, as in addition we usually need tens of or more energy
points, the whole process will be very time-consuming. In MS theory, we have
observed a fact that the contribution of high angular momenta is much smaller
than that of low angular momenta and can be considered as a perturbation.
Therefore, we expect the contribution of these high angular momenta to be
included efficiently in some approximative ways without loss of
accuracy.\par

Zhang $et\ al.$ \cite{zhang1992multiple} suggested an approximate method where
the phase shifts of high angular momenta ($l_{pt}\,<\,l\,\leq\,l_{max}$) are
neglected, so that the computation is considerably simplified. They obtained a
good normalization of the wavefunction with small $l_{pt}$. Applications and
more discussions are given by R. Zeller \cite{zeller2013projection} and A. Alam
$et\ al.$ \cite{alam2014green}.\par

However, it was found that when Muffin-Tin (MT) approximation does not work or when
angular momentum convergence is slow, Zhang's method may not be so accurate, so that it
becomes necessary to go beyond. In our formalism, while keeping
to solve single-site problem with $l_{max}$, we introduce approximations
directly to the elements of the scattering matrix $I-tg$: some unimportant
elements are neglected, so that some submatrices of $I-tg$ become sparse.
While our formalism loses no or a little efficiency, the accuracy is improved considerably
 compared to Zhang's method.\par

\section{Formalism}

\subsection{Two forms of Green's function}
\label{sec_GF}

Our formalism is based on the Full-Potential Multiple Scattering (FPMS) theory
with space-filling cells developed by K. Hatada $et\ al.$
\cite{FP_Hatada,FP_Hatada2,FP_Hatada3}. Here, space is partitioned by
nonoverlapping space-filling cells or Voronoi polyhedra. When a Voronoi
polyhedron does not contain any atom or is in the interstitial region but still
contains charge density, it is called an empty cell (EC). The local
Schr\"odinger equation is solved without the limit of the geometrical shape of
the potential, since the potential is not expanded in spherical harmonics.\par
In MS theory, two forms of Green's function are commonly used:
\begin{equation}
\centering
G(\vec{r_i},\vec{r'_j};E) = \sum_{LL'}\bar{\Phi}_L^i(\vec{r_i})([I-tg]^
{-1}t)_{LL'}^{ij}\bar{\Phi}^j_{L'}(\vec{r'_j})-\delta_{ij}\sum_L\bar
{\Phi}_L^i(\vec{r_<})\Lambda_L^i(\vec{r_>}) \label{eq:GF1}
\end{equation}
and
\begin{eqnarray}
\centering
G(\vec{r_i},\vec{r'_j};E) = \sum_{LL'}\underline{\Phi}_L^i(\vec{r_i})(g[I-tg]^
{-1})_{LL'}^{ij}\underline{\Phi}^j_{L'}(\vec{r'_j})-\delta_{ij}\sum_L\underline
{\Phi}_L^i(\vec{r_<})\Psi_L^i(\vec{r_>}), \label{eq:GF2}
\end{eqnarray}
where $\vec{r_i}$ is the coordinate with respect to the center of scattering
site $i$. $\Psi$ and $\Lambda$ are the irregular solutions of local
Schr\"odinger equation which match smoothly to spherical Hankel and Bessel
functions, respectively, at the cell boundary. $r_< (r_>)$ is the smaller
(larger) of $r$ and $r'$. The definitions of $\underline{\Phi}$ and
$\bar{\Phi}$ are:
\begin{eqnarray}
\bar{\Phi}_L \equiv \sum_{L'} S_{L'L}^{-1}\Phi_{L'} \label{eq:phibar}
\end{eqnarray}
and
\begin{eqnarray}
\underline{\Phi}_L \equiv \sum_{L'} E_{L'L}^{-1}\Phi_{L'},
\end{eqnarray}
where $\Phi$ is the solution of local Schr\"odinger equation which behaves as
the spherical Bessel function of the first kind at the origin. The definitions
of $E$ and $S$ matrices will be given in Appendix and they are computed using
values on the surface of the bounding sphere of the cell.\par The transition
operators are defined as
\begin{eqnarray}
t_{LL'}=-\sum_{L''} S_{LL''}(E^{-1})_{L''L'} \label{eq:t}.
\end{eqnarray}
The first form of Green's function in Eq. \ref{eq:GF1} is more efficient than
the second form in Eq. \ref{eq:GF2}, since multiplying matrix $[I-tg]^{-1}$ by
site-diagonal matrix $t$ is much easier than multiplying $[I-tg]^{-1}$ by $g$.
Actually, in MS theory, usually $[I-tg]^{-1}t$ is defined as matrix $\tau$. On
the other hand, the second form of Green's function in Eq. \ref{eq:GF2} is
numerically a bit more stable, since $S$ matrix and $\Lambda$ functions have
poles on the real energy axis. Through the end of this article, we will only
apply the first form.

\subsection{Matrix partitioning and approximations for matrix inversion}
\label{subsec_matp}

We define matrices $M\,=\,I-tg$ and $\mathscr{M}\,=\,[I-tg]^{-1}$. These
matrices are truncated at $l_{max}$. If we partition $M$ by a particular
orbital angular momentum $l_{pt}$ into four submatrices, we have
\begin{equation}
M=
\left[{\begin{array}{cc}
M_{ss'} & M_{sb'} \\ M_{bs'} & M_{bb'}
\end{array}}\right]=
\left[{\begin{array}{cc}
A & B \\ C & D
\end{array}}\right],
\end{equation}
where the subscript $s$ ($s'$) and $b$ ($b'$) correspond respectively to $l$ small
($0<l\leq{l_{pt}}$) and large ($l_{pt}<l<l_{max}$). The submatrices
are:
\begin{eqnarray}
A = I-t_{sf}g_{fs'}, \\
B = -t_{sf}g_{fb'}, \label{eq:B}\\
C = -t_{bf}g_{fs'}, \label{eq:C}\\
D = I-t_{bf}g_{fb'},
\end{eqnarray}
where the subscript $f$ indicates that the full angular momenta are included. The
Einstein summation convention is used, and the summations are only on two
neighboring subscripts.\par We define also
\begin{equation}
\mathscr{M}=
\left[{\begin{array}{cc}
\mathscr{M}_{ss'} & \mathscr{M}_{sb'} \\ \mathscr{M}_{bs'} & \mathscr{M}_{bb'}
\end{array}}\right]=
\left[{\begin{array}{cc}
\mathscr{A} & \mathscr{B} \\ \mathscr{C} & \mathscr{D}
\end{array}}\right].
\end{equation}
Therefore, we have the relations:
\begin{eqnarray}
\mathscr{A} = [ A - B D^{-1} C ]^{-1}, \label{eq:tildeA}\\
\mathscr{B} = -\mathscr{A}BD^{-1}, \label{eq:tildeB}\\
\mathscr{C} = -D^{-1}C\mathscr{A}, \label{eq:tildeC}\\
\mathscr{D} = D^{-1} + D^{-1}C\mathscr{A}BD^{-1}. \label{eq:tildeD}
\end{eqnarray}
Usually matrix $A$ is very dense. Except for the diagonal elements, the
elements of matrix $D$ are very small, so that we approximate $D$ and $D^{-1}$
by the unit matrix. This approximation works well when we study the energy
region below or several tens of electron volt (eV) above the Fermi level.
Afterwards, we have the approximate relations:
\begin{eqnarray}
\mathscr{A} = [ A - B C ]^{-1}, \label{eq:Minvss}\\
\mathscr{B} = - \mathscr{A} B, \label{eq:Minvsb}\\
\mathscr{C} = - C \mathscr{A}, \label{eq:Minvbs}\\
\mathscr{D} = I + C \mathscr{A} B. \label{eq:Minvbb}
\end{eqnarray}
In this work, matrix $D$ is always approximated by the unit matrix, i.e., we
will use Eq. \ref{eq:Minvss} - \ref{eq:Minvbb} instead of Eq. \ref{eq:tildeA} -
\ref{eq:tildeD}.

A further approximation is to treat $B$ as a sparse matrix. In Sec.
\ref{sec_results}, we prove that this approximation works well. The way to
construct an approximate matrix of $B$ is straightforward: a particular
percentage, e.g., 1\%, of elements, whose absolute values are larger than the
rest are kept in matrix $B$  while all the others are set to zero. The
computation time of sparsification is then proportional to the size of the
matrix $B$. Furthermore, the fact, that the threshold absolute value used to
make matrix $B$ sparse varies slowly with energy, makes this computation even
cheaper. Whether $C$ can be approximated by a sparse matrix for computing
$\mathscr{C}$ and $\mathscr{D}$ needs further studies, but we have checked that
it is not satisfactory to treat $C$ as a sparse matrix for computing
$\mathscr{A}$.\par

\subsection{Comparisons with Zhang's method}
\label{subsec_zhang}

The original Zhang's method uses the second form of Green's function (Eq.
\ref{eq:GF2}) and neglects the phase shifts of high angular momenta. It can be
seen as the combination of two approximations: (i) The spherical-wave basis is
truncated by $l_{pt}$, so that when $l_{pt}\,<\,l\,\leq\,l_{max}$, $\Phi_L$ and
$\Psi_L$ are respectively the spherical Bessel and Hankel functions. (ii) For
matrix $I-tg$, $t_{LL'}$ with $l>l_{pt}$ or $l'>l_{pt}$ is approximated by
zero, so that, $B=-t_{ss'}g_{s'b}$ (Eq. \ref{eq:C}) and $C=0$ (Eq. \ref{eq:B}).\par

Zhang's method works well when MT approximation is valid and angular momentum
convergence is fast. However, it may be not suitable when MT approximation
behaves poorly or breaks down because of the following three points:\par

(i) $\Phi_L=\sum_{L'}R_{L'L}Y_{L'}$ and $\Phi_L$ with $l<l_{pt}$ is now strongly
affected when increasing $l_{pt}$, while in MT approximation, $\Phi_L=R_lY_L$
with $l<l_{pt}$ is not affected by the chosen of $l_{pt}$.\par

(ii) The absolute value of $t_{LL'}$ with $l$ or $l'$ $>$ $l_{pt}$ may not be
small now. On the other hand, in MT approximation, $t_{LL'}=\delta_{LL'}t_l$ so
that the off-diagonal terms are zero, and $t_l$ with $l>l_{pt}$ is usually
quite small if $l_{pt}$ is not small. This point may strongly break the second
approximation of Zhang's method. In the application of X-ray absorption
spectroscopy (XAS) of graphene, we have observed that the absolute value of
$t_{LL'}$ with $l>l_{pt}$ and $l'\leq l_{pt}$ can be quite large.\par

(iii) The angular momentum convergence may be slower now, if the potential is highly
anisotropic. By contrast, in MT approximation, this convergence is mainly controlled
by the energy, so that when we study the low energy region, a standard
calculation with $l_{max}=$2 or 3 can already give reasonable results for some
physical properties.

It should be noticed that in the following discussions, when we mention Zhang's
method, it always contains only the second approximation.

\subsection{Formalism in different types of calculations}
\label{subsec_complexity}

In many studies, the full information contained in Green's function is not
necessary, that is to say, only parts of elements of matrix $\tau$ defined in Sec.
\ref{sec_GF} are needed. For the following discussion, we define the number of
scattering sites as $N_{sc}$, and the dimension of matrix $A$ and $D$ as
$a=N_{sc}(l_{pt}+1)^2$ and $b=N_{sc}[(l_{max}+1)^2-(l_{pt}+1)^2]$ respectively.
Moreover, in this work, matrix $D$ is always approximated by the unit matrix,
i.e., Eq. \ref{eq:Minvss} - \ref{eq:Minvbb} are applied. There are many methods
to compute the full or partial matrix inversion, In this work, we only use LU
decomposition.\par

\subsubsection{$\tau^{00}$ or $\tau^{i0}$}

Here, "$0$" represents one particular scattering site while "$i$" can be any
site. For core-level spectroscopies, only $\tau^{00}$ is needed and "$0$" is
restricted to the absorber where a core electron is excited. For photoemission,
what we need is $\tau^{i0}$.\par $\tau^{i0}$ can be computed in the following
steps:\par 1. Compute $BC$. The number of multiplications
 $N_{op}$ is $c_spa^2b$, where $p$ is the percentage of nonzero
elements of $B$. As $l_{pt}$ and $N_{sc}$ increase, we can expect that $p$ can
be suitably chosen as a smaller value. $c_s$ ($>$1) is introduced to consider
the fact that the implementation of matrix multiplication of a sparse matrix by
a dense matrix is often not well performed so that the efficiency is less than
expected, or needs additional operations other than the necessary
multiplications and summations of two values, although $c_s$ does not really
affect the number of multiplications. In this work, the normal dense matrix
multiplication is done using the LAPACK math library, while a sparse matrix
being multiplied by a dense matrix is done without the use of any math library.
Therefore, $c_s$, in this work, is about 3.\par 2. Compute $\mathscr{A}^{j0}$.
By LU decomposition, we obtain $A-BC=LU$, where $L$ and $U$ are respectively a
lower and a upper triangular matrix. The computation is dominated by LU
decomposition, so that $N_{op}\,\approx\,a^3/3$.\par 3. Compute
$\mathscr{B}^{j0}$. We just need to solve $(l_{max}+1)^2-(l_{pt}+1)^2$ linear
equations, so that the complexity of this process is $O(a^2)$.\par 4. Compute
$\mathscr{C}^{j0}$. The complexity of this process is $O(ab)$.\par 5. Compute
$\mathscr{D}^{j0}$. The complexity of this process is $O(ab)$.\par 6. Compute
$\tau^{i0}=\mathscr{M}^{i0}t^0$. The complexity of this process is
$O(a+b)$.\par The whole computation is dominated by the first two steps, so
that $N_{op}\,\approx\,a^3/3+c_spa^2b$, if $N_{sc}$ is not small. For the
computation of $\tau^{00}$, $N_{op}$ is smaller but still approximated by
$a^3/3+c_spa^2b$, if $N_{sc}$ is not small. In Zhang's method, since $C=0$,
$N_{op}\,\approx\,a^3/3$. In standard calculations, the computation time is
dominated by the LU decomposition of the full $M$ matrix, so that
$N_{op}=(a+b)^3/3$. If $l_{max}$\,=\,6, $l_{pt}$\,=\,3, $p$\,=\,1\% and
$c_s$\,=\,3, $N_{op}$ of our method is about 4.1\% of $N_{op}$ of the standard
calculation.

\subsubsection{$\tau^{ii}$}

For the calculations of the density of states (DOS) and related quantities,
e.g., the total energy, only the site-diagonal elements of matrix $\tau$, i.e.,
$\tau^{ii}$, are needed. $\tau^{ii}$ can be computed in the following
steps:\par 1. Compute $BC$. $N_{op}=c_spa^2b$.\par 2. Compute
$\mathscr{A}^{ii}$. $N_{op}=a^3$.\par 3. Compute $\mathscr{B}^{ii}$. the
complexity of this process is $O(N_{sc}^2)$.\par 4. Compute $\mathscr{C}^{ii}$.
From Eq. \ref{eq:Minvbs}, the complexity of this process is $O(N_{sc}^2)$.\par
5. Compute $\mathscr{D}^{ii}$. From Eq. \ref{eq:Minvsb} and \ref{eq:Minvbb},
the complexity of this step is $O(N_{sc}^2)$.\par 6. Compute
$\tau^{ii}=\mathscr{M}^{ii}t^i$. The complexity is $O(N_{sc})$.\par The total
number of multiplications is $N_{op}\,\approx\,a^3+c_spa^2b$, while in standard
calculations, $N_{op}=(a+b)^3$. In Zhang's method, $N_{op}\,\approx\,a^3+a^2b$.
If $l_{max}$\,=\,6, $l_{pt}$\,=\,3, $p$\,=\,1\% and $c_s$\,=\,3, $N_{op}$ in
our method is about 3.7\% of $N_{op}$ of the standard calculation.

\subsubsection{Density of states}

The total DOS is
\begin{eqnarray}
n(E)=\sum_i n_i(E)+n_I(E),
\end{eqnarray}
where $n_I(E)$ is the DOS of the interstitial charge. If space is filled by Voronoi
polyhedra, $n_I(E)$ will reduce to zero. $n_i(E)$ is the integral DOS in the
cell $i$,
\begin{eqnarray}
n_i(E)=-\frac{1}{\pi}Im\int_{\Omega_i} G(\vec{r},\vec{r},E)d\vec{r}.
\end{eqnarray}
We substitute $\Phi_L(\vec{r})=\sum_{L'} R_{L'L}(r)Y_{L'}(\hat{\vec{r}})$,
$\Lambda_L(\vec{r})=\sum_{L'} \bar{R}_{L'L}(r)Y_{L'}(\hat{\vec{r}})$, where
$Y_L$ are the real spherical harmonics, Eq. \ref{eq:phibar} and Eq.
\ref{eq:GF1} into the above equation. We obtain
\begin{eqnarray}
n_i(E) = -\frac{1}{\pi}Im \left [\sum_{LL'} \Big ( \tau^{ii}_{LL'}\sum_{L_1L_2}
S^{-1}_{L_1L} S^{-1}_{L_2L'} \rho_{L_1L_2} \Big ) + \sum_{L'L} S^{-1}_{L'L}
\bar{\rho}_{L'L} \right ], \\
\rho_{L_1L_2} = \sum_{L_3L_4} \int r^2drR_{L_3L_1}(r)R_{L_4L_2}(r) \int
d\hat{\vec{r}} Y_{L_3}(\hat{\vec{r}})Y_{L_4}(\hat{\vec{r}})\theta(\vec{r}), \\
\bar{\rho}_{L'L} = \sum_{L_3L_4} \int r^2dr\bar{R}_{L_3L'}(r)R_{L_4L}(r) \int
d\hat{\vec{r}}Y_{L_3}(\hat{\vec{r}})Y_{L_4}(\hat{\vec{r}})\theta(\vec{r}),
\end{eqnarray}
where $\theta(\vec{r})$ is the shape function which vanishes outside the cell.
For the sake of simplicity, we only calculate in this work the local DOS $n^{in}_i(E)$
where the integration region is limited to the inscribed sphere of a cell, so
that
\begin{eqnarray}
\int d\hat{\vec{r}}Y_{L_3}(\hat{\vec{r}})Y_{L_4}(\hat{\vec{r}})\theta(\vec{r})=
\delta_{L_3L_4},
\end{eqnarray}
and
\begin{eqnarray}
\rho_{L_1L_2}=\sum_{L_3} \int_0^{R^{in}} r^2drR_{L_3L_1}(r)R_{L_3L_2}(r), \\
\bar{\rho}_{L'L}=\sum_{L_3} \int_0^{R^{in}} r^2dr\bar{R}_{L_3L'}(r)R_{L_3L}(r),
\end{eqnarray}
where $R^{in}$ is the radius of the inscribed sphere. If $r\,>\,R^{in}$, we can
use the relation
\begin{eqnarray}
\int d\hat{\vec{r}}Y_{L_3}(\hat{\vec{r}})Y_{L_4}(\hat{\vec{r}})=\sum_{L_5} C(L_5
L_3|L_4) Y_{L_5}(\hat{\vec{r}}),
\end{eqnarray}
where $C(L_5 L_3|L_4)$ is the Gaunt coefficient.

\section{Results and discussions}
\label{sec_results}

\subsection{DOS of Cu crystal}

\begin{figure}[t]
\includegraphics[width=0.8\columnwidth]{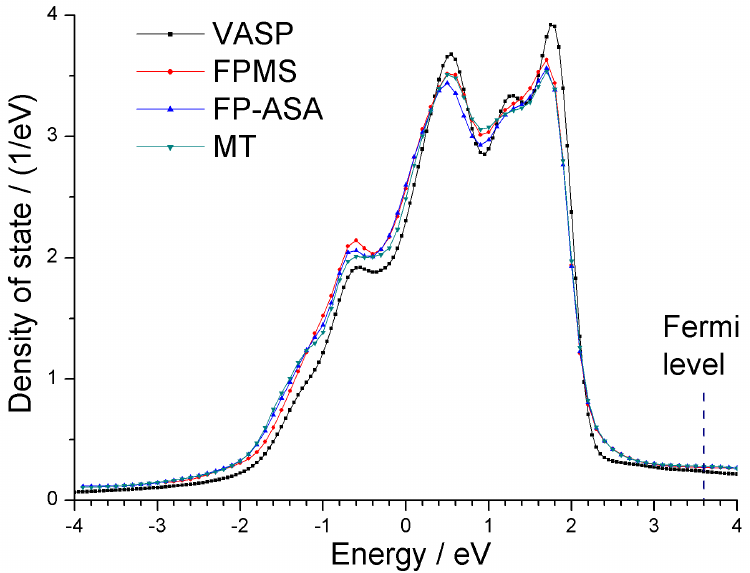}
\caption{\label{dos_Cu_comp}Calculated local DOS of a Cu sphere of radius
1.276\ \AA\ in fcc Cu crystal using different methods. VASP corresponds to the
VASP code with PAW (Projector Augmented Wave) method. MT, FP-ASA and FPMS represent three kinds of Multiple Scattering calculations.
"MT" uses Muffin-Tin approximation, while "FP-ASA" uses atomic-sphere approximation but inside each atomic sphere, potential is not approximated to sphereical.
In "FPMS", no approximation is applied to the shape of potential and  space is partitioned by
nonoverlapping space-filling cells or Voronoi polyhedra. For all MS calculations,
a cluster of radius 11\ \AA, containing 459 atoms, is used.}
\end{figure}

For the standard test case, we first study a Cu crystal, for which it is well known that
MT approximation is a good approximation. Therefore, we expect Zhang's
method (see Sec. \ref{subsec_zhang}) to work well. However, we will show that the
accuracy is greatly improved by our method without any loss of efficiency (see
the discussion of computation time in Sec. \ref{subsec_complexity}).\par

In Fig. \ref{dos_Cu_comp}, we compare the local DOS of a Cu sphere of radius
1.276\ \AA\ in fcc Cu crystal computed with different methods. To make the
comparison easier, a 0.15 eV Gaussian broadening is used. VASP corresponds to the
VASP code with PAW (Projector Augmented Wave) method. MT, FP-ASA and FPMS represent three kinds of Multiple Scattering calculations.
"MT" uses Muffin-Tin approximation, while "FP-ASA" uses atomic-sphere approximation but inside each atomic sphere, potential is not approximated to sphereical.
In "FPMS", no approximation is applied to the shape of potential and  space is partitioned by
nonoverlapping space-filling cells or Voronoi polyhedra. In our FP-ASA and MT calculations, the diameters of the spheres
are taken to be 10\% larger than the nearest-neighbor distance. In the FPMS calculation, we have
added an empty cell on the center of the fcc cube of the Cu crystal in order to
fill the space.\par

In VASP calculation, the plane-wave cut-off energy is 400 eV and a
9$\times$9$\times$9 and 21$\times$21$\times$21 Monkhorst-Pack K-point sampling
were used respectively to generate the charge density and compute the DOS. The exchange-correlation functional ($V_{xc}$) proposed by Ceperley and Alder
\cite{CA} and parameterized by Perdew and Zunder \cite{PZ}, named CA-PZ, has been employed.\par

\begin{figure}[t]
\subfigure{
\label{dos1_Cu_lptt}
\includegraphics[width=0.5\columnwidth]{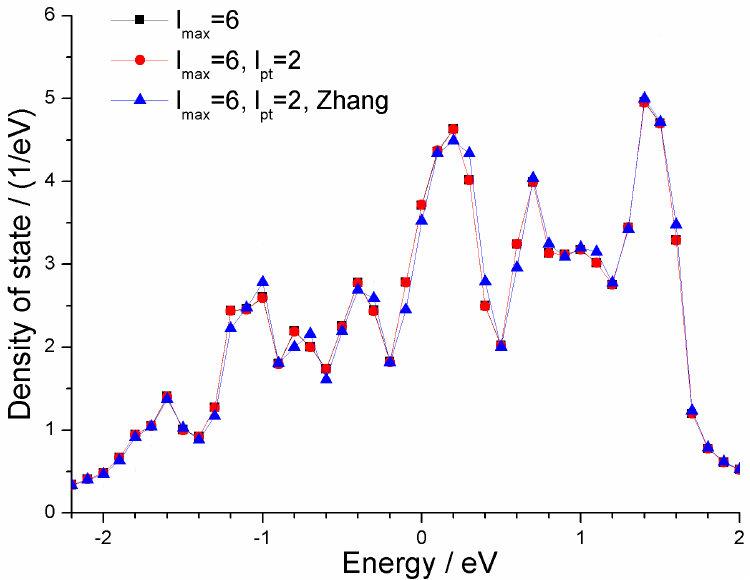}
}
\subfigure{
\label{dos2_Cu_lptt}
\includegraphics[width=0.5\columnwidth]{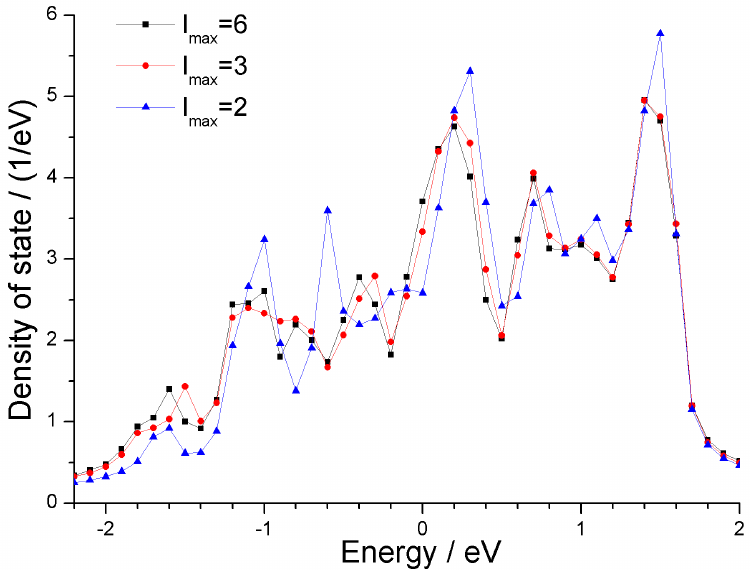}
}
\caption{\label{dos_Cu_lptt}Calculated local DOS on the central Cu atom of a
cluster, of radius 8\ \AA, containing 177 atoms, of fcc Cu crystal by the
FPMS method. No additional Gaussian broadening is used.}
\end{figure}

\begin{figure}[t]
\subfigure{
\label{dev1_Cu_lptt}
\includegraphics[width=0.5\columnwidth]{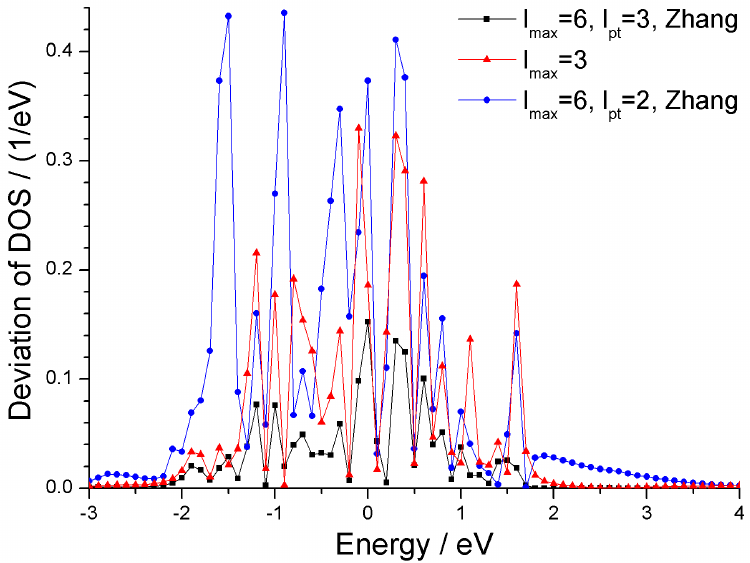}
}
\subfigure{
\label{dev2_Cu_lptt}
\includegraphics[width=0.5\columnwidth]{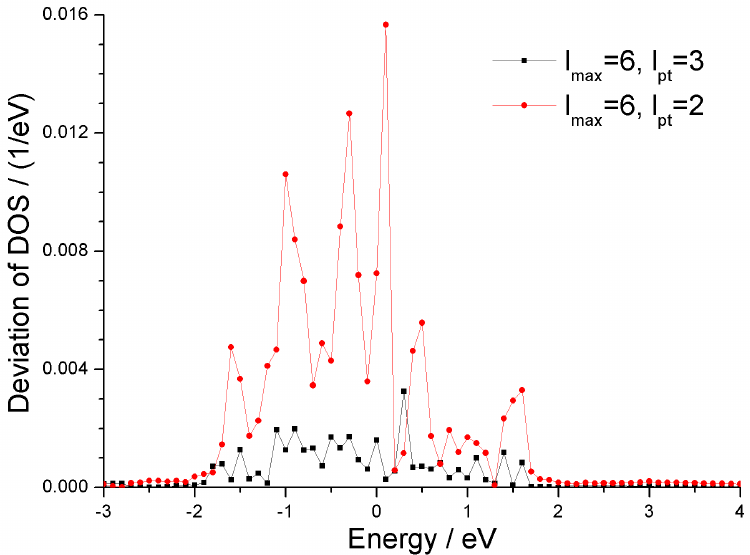}
}
\caption{\label{dev_Cu_lptt} The absolute values of the deviations between the
results of different FPMS calculations and the the result of the standard FPMS
calculation with $l_{max}=6$. The magnitude of the values in the right panel is
much smaller than that in the left panel.}
\end{figure}

In Fig. \ref{dos_Cu_comp}, in order to reproduce the bulk properties (so as to compare with VASP result) by our real-space MS calculations,
in all three MS calculations, we use a huge cluster of radius 11\ \AA, containing 459 atoms. Morever, for the optical potential, its real part is
taken as CA-PZ $V_{xc}$, while its imaginary part is a small constant. By contrast, in Fig. \ref{dos_Cu_lptt} and \ref{dev_Cu_lptt}, since our purpose is to check the new algorithm of matrix inversion, the radius of the cluster is chosen smaller as 8\ \AA, containing 177 atoms. Additionally, the real part of the optical potential is the
Hedin-Lundqvist (HL) potential \cite{HL}. As we have checked, the MS results using CA-PZ $V_{xc}$ and HL potential are different by
mainly a small energy shift. The self-consistent charge density and
electrostatic potential for MS calculation is obtained from the ES2MS package
\cite{ES2MS} from an all-electron charge density and a pseudo electrostatic
potential generated by the VASP code. We treat $l = 6$ as the converged value
of $l_{max}$. However for simplicity, in Fig. \ref{dos_Cu_comp}, $l_{max}$ is
set to 5. We checked that the differences between the result using $l_{max}=5$
and that using $l_{max}=6$ is very small.

From Fig. \ref{dos_Cu_comp}, we see that all three MS results are similar to
that by VASP code. The differences between MS and VASP results can be
attributed to the following points: (i) A larger radius of the cluster may be
needed for the better convergence of the MS calculation. (ii) The local DOS of
Cu by VASP is the sum of s-, p- and d-electron DOS inside a sphere, since the
VASP code usually does not give other components in the output files. (iii) MS
uses an energy-dependent basis while the VASP calculation uses an
energy-independent basis.

From Fig. \ref{dos_Cu_lptt}, we can see the angular momentum convergence of the DOS
of the Cu crystal is quite fast, e.g., the FPMS result with $l_{max}=3$ is close to
that with $l_{max}=6$. With $l_{max}=6$ and $l_{pt}=2$, the result by our
method is nearly the same as the result of the standard calculation with
$l_{max}=6$, while the result by Zhang's method is worse but better than the
$l_{max}=3$ result.\par

Fig. \ref{dev_Cu_lptt} shows the absolute values of the deviations between results
of different FPMS calculations and the result of the standard FPMS calculation
with $l_{max}=6$. The accuracy of our method is tens of or one hundred times
better than Zhang's method.

\begin{figure}[t]
\includegraphics[width=0.8\columnwidth]{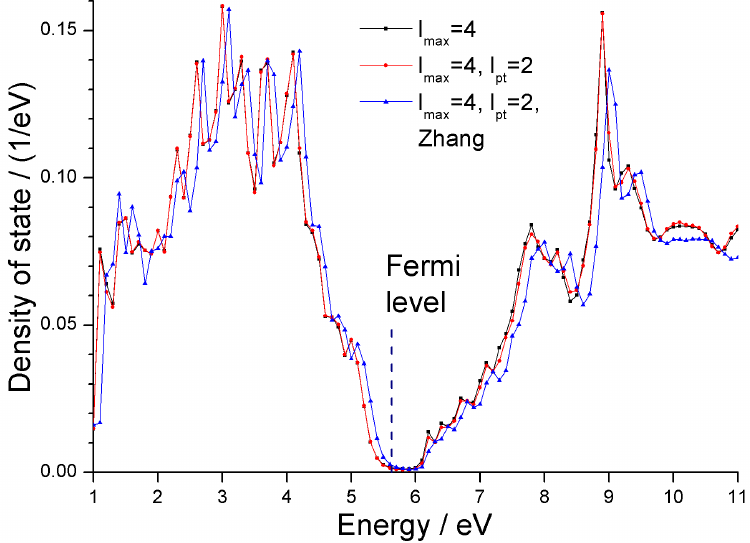}
\caption{\label{dos_Si_lptt}Calculated local projected DOS (T$_2$ irreducible
representation) of the central Si atomic sphere, of radius 1.175\ \AA, of a
cluster of silicon by FPMS. The cluster radius is 24\ \AA, containing 2917
atoms. The energy gap calculated by our method, with $l_{max}=4$ and
$l_{pt}=2$, is very close to the gap calculated by VASP code, while the gap
given by Zhang's method is much worse. However, since LDA is applied to $V_{xc}$ or optical potential,
the energy gaps given by VASP calculation and our method are not very close to the experimental one.}
\end{figure}

\subsection{DOS of silicon}

As discussed in Sec. \ref{subsec_zhang}, when FP becomes important, the errors
in Zhang's method may be large. In open systems, especially when covalent bonds
are present, MT approximation may fail. Moreover, from our experience, as the
element studied becomes lighter, FP effects may be stronger. In Ref.
\cite{FP_Hatada3}, it was found that FP is necessary for $\alpha$-quartz.
Therefore, in order to study silicon, FP is expected to be necessary. In Ref.
\cite{FP_GE_Xu}, the authors have shown that FP effects are strong for
graphene-related systems. In the following two subsections, we will study
silicon and graphene systems where FP effects are large, and compare the
results by Zhang's and by our method.\par

In order to reproduce the local DOS of the Si atom in silicon, a very large
cluster may be needed for real-space methods \cite{luo2008carrier}. For
simplicity, we use a cluster of radius 24\ \AA, containing 2917 atoms.
Moreover, we employe the T$_d$ point group symmetry and focus only on the T$_2$
irreducible representation so that we obtain the local projected DOS where the
wavefunction is projected onto the basis $(x,y,z)$. Since
the ground-state properties of silicon are dominated by $p$-electron orbitals,
the local projected DOS obtained has a similar shape as the total DOS of
silicon and can reproduce the same energy gap.\par

We have checked that when we use a cluster of radius 18\ \AA, corresponding to
1207 Si atoms, the local DOS of the central Si atom obtained by FPMS is quite
similar to the local DOS of the Si atom in silicon calculated with the VASP
code, except that the DOS by FPMS at Fermi level is about 0.005 state per
electron volt while it is much smaller when computed with the VASP code. Since
we use the self-consistent crystal potential in our real-space cluster
calculations, this potential is not suitable for scattering sites near the
boundary of the cluster. While the DOS near the Fermi level is very sensitive
to the accuracy of the potential, our real-space method is not suitable to
study the ground-state properties of semiconductor crystals and nanocrystals.
To solve this problem, the reciprocal-space version of our FPMS method and the
self-consistent cluster potential are needed. Additionally, the truncated
crystal approximation may be useful for constructing the cluster potential from
a bulk calculation \cite{zhang1993electronic,franceschetti1996gaas}.

\begin{figure}[t]
\includegraphics[width=1.0\columnwidth]{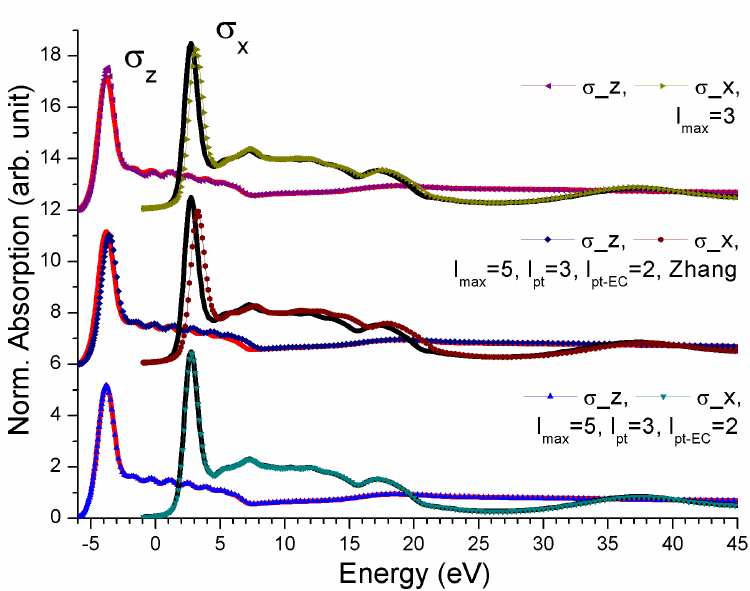}
\caption{\label{G_xas_lptt}Calculated XAS of a graphene cluster of radius 20\
\AA, containing 481 Carbon atoms and 1902 empty cells. Three calculations are
compared with the standard calculation with $l_{max} =5$. The red (black) solid
line represents $\sigma_z(x)$ of XAS of graphene by the standard calculation
with $l_{max}=5$. "$l_{pt}=3, l_{pt-EC}=2$" correspond respectively to $l_{pt}$
of the atomic cells set to 3 and $l_{pt}$ of the empty cells set to 2. "Zhang"
corresponds to the Zhang's method (see Sec. \ref{subsec_zhang}).
$\sigma_{z(x)}$ is polarized absorption cross section with electric field along
z(x) axis.}
\end{figure}

In Fig. \ref{dos_Si_lptt}, we compare the local projected DOS (T$_2$
irreducible representation) of the central Si atomic sphere, of radius 1.175\
\AA, of a cluster of silicon by the standard FPMS calculation with $l_{max}=4$,
by our approximate method with $l_{max}=4$ and $l_{pt}=2$ and by Zhang's method
with $l_{max}=4$ and $l_{pt}=2$. Empty cells are added at the positions
described in Ref. \cite{der1995self}. The errors on the result by our method is
much smaller than those by Zhang's method.

\subsection{XAS of graphene}

In Ref. \cite{FP_GE_Xu}, we have applied the real-space FPMS method with
space-filling cells with self-consistent potential to study XAS of graphene and
graphene oxide. The fact that good agreement with experiments was always
achieved demonstrated the accuracy of our FPMS method. For simplicity, in this
work, we focus on XAS of a graphene cluster of radius 20\ \AA, containing 481
Carbon atoms and 1902 empty cells. Computation details are given in Ref.
\cite{FP_GE_Xu}.

In Fig. \ref{G_xas_lptt}, we find that with our method, if we set
$l_{max}=5$, $l_{pt}$ of atomic and empty cells respectively as 3 and 2, the
results are nearly the same as the results obtained  by standard calculations with
$l_{max}=5$. Since the number of empty cells is several times more than the
number of atomic cells, the efficiency of this calculation is close to the case
where $l_{pt}$ of all scattering sites is 2. Moreover, the results by Zhang's
method are worse than the results of the standard calculations with $l_{max}=3$.

\section{conclusion}

We have presented an efficient Multiple Scattering formalism based on the
partitioning of the scattering matrix $I-tg$ by a particular orbital angular
momentum $l_{pt}$. By introducing approximations to the submatrices of $I-tg$,
the computation of the matrix inversion of $I-tg$ and Green's function are
simplified a lot with a very little loss of accuracy. We have discussed the detailed
behaviour of our method for different types of calculations where only parts of
the matrix elements of $\tau$ are needed. We have applied our method to
calculate the local DOS of the central atomic sphere of fcc a Cu crystal and a
silicon one, and the XAS of graphene. We have found that with small $l_{pt}$,
the results with our method are nearly the same as the results of the standard
calculations with large $l_{max}$. With the same values of $l_{max}$ and
$l_{pt}$, the accuracy of our method is much higher than that of Zhang's method
where more approximations are introduced. Moreover, we have found that in the
studies of graphene and silicon, the results of Zhang's method are not
satisfactory. A possible reason is that the MT approximation breaks down and
the phase shifts of high angular momenta are not negligible anymore.

\section*{Appendix}

In Sec. \ref{sec_GF}, we have introduced basis functions $\Phi_L$ and it can be
expanded as $\Phi_L(\vec{r}) = \sum_{L'}R_{L'L}(r)Y_{L'}(\hat{r})$, where
$Y_{L'}(\hat{r})$ are real spherical harmonics. With the radial wavefunction
$R_{L'L}(r)$, we can compute\par

\setcounter{equation}{0}
\renewcommand{\theequation}{A.\arabic{equation}}
\begin{eqnarray}
E_{LL'}=(R_b)^2W[-i\kappa h^+_l,R_{LL'}]
\end{eqnarray}
and
\begin{eqnarray}
S_{LL'}=(R_b)^2W[j_l,R_{LL'}],
\end{eqnarray}
where $R_b$ is the radius of the bounding sphere of the scattering cell. $j_l$
and $h^+_l$ denote spherical Bessel and Hankel functions of order $l$,
respectively. The Wronskians $W[f,g]=fg'-gf'$ are calculated at $R_b$. $\kappa$
is the electronic momentum relative to the constant interstitial potential.

\section*{Acknowledgements}

K. H. acknowledges a funding of the European FP7 MS-BEEM (Grant Agreement No.
PIEF-GA-2013-625388). Parts of this work have been funded by European FP7
MSNano network under Grant Agreement No. PIRSES-GA-2012-317554, by COST Action
MP1306 EUSpec, by the National Natural Science Foundation of China (U1232131,
11375198), by the Science Fund for Creative Research Groups of the NSFC
(11321503) and by JSPS KAKENHI Grant Number 25887008.

\section*{References}

\bibliographystyle{unsrt}
\bibliography{ref_lptt}

\end{document}